\documentclass[]{emulateapj}

\usepackage{natbib}

\usepackage{amsmath}
\usepackage{dcolumn,multirow,ccaption}
\usepackage{graphicx}
\usepackage{color}
\usepackage{ccaption}

\usepackage{natbib}

\usepackage{amsmath,amsthm,amssymb,subfigure}
\slugcomment{Not to appear in Nonlearned J., 45.}

\shorttitle{Star formation efficiency in the Barred Spiral Galaxy \object{NGC 4303}}
\shortauthors{Momose et al.}

\begin{document}

\title{Star formation efficiency in the Barred Spiral Galaxy \object{NGC 4303}}

\author{Rieko Momose\altaffilmark{1,2}, Sachiko K. Okumura\altaffilmark{2}, Jin Koda\altaffilmark{3,4}, Tsuyoshi Sawada\altaffilmark{2}}

\altaffiltext{1}{Department of Astronomy, University of Tokyo, Hongo, Bunkyo-ku, Tokyo 113-0033, Japan; momo.s.rieko@nao.ac.jp}
\altaffiltext{2}{National Astronomical Observatory of Japan, Mitaka, Tokyo 181-8588, Japan; sokumura@nro.nao.ac.jp, tsawada@alma.cl}
\altaffiltext{3}{Department of Physics and Astronomy, Stony Brook University, Stony Brook, NY 11794-3800, USA; Jin.Koda@stonybrook.edu}
\altaffiltext{4}{California Institute of Technology, MS 105-24, Pasadena, CA 91125, USA}

\begin{abstract}
We present new $^{12}$CO(J=1-0) observations of the barred galaxy NGC 4303 using the Nobeyama 45m telescope (NRO45) and the Combined Array for Research in Millimeter-wave Astronomy (CARMA). The H$\alpha$ images of barred spiral galaxies often show active star formation in spiral arms, but less so in bars. We quantify the difference by measuring star formation rate and efficiency at a scale where local star formation is spatially resolved. Our CO map covers the central 2$\farcm$3 region of the galaxy; the combination of NRO45 and CARMA provides a high fidelity image, enabling accurate measurements of molecular gas surface density. We find that star formation rate and efficiency are twice as high in the spiral arms as in the bar. We discuss this difference in the context of the Kennicutt-Schimidt (KS) law, which indicates a constant star formation rate at a given gas surface density. The KS law breaks down at our native resolution ($\sim$ 250 pc), and substantial smoothing (to 500 pc) is necessary to reproduce the KS law, although with greater scatter.
\end{abstract}


\section{INTRODUCTION}
Barred galaxies often reveal a striking difference in star formation activity between the bar and outer spiral arms(e.g. \citealp{dow96, she02}). Most star forming regions (H$\alpha$ emission) are associated with the outer spiral arms rather than with the bar, even when both host a similar amount of molecular gas.

The difference is often attributed to strong shear motions along the bar, by which giant molecular clouds (GMCs) could be pulled apart and disrupted before the ignition of star formation (e.g. \citealp{dow96, shci02}). On the other hand, according to \citet{kod06}, the observed velocity gradient in a bar is not large enough to shred gravitationally-bound GMCs.

Unfortunately, previous studies suffered from large uncertainties in measurement of gas surface density, due to the poor image quality (fidelity) of interferometer observations, or to low sensitivity and resolution even when the observations from a single dish telescope are included. Interferometers have the intrinsic missing flux problem, and cannot measure the gas surface density accurately. Most interferometric studies covered only a central bar region including the nucleus, and did not extend to outer spiral arms, because of the limited field-of-view. They studies concentrated on nuclear gas condensations and the gas fueling galactic centers(e.g. \citealp{sak99a}), and did not contrast the difference between the bar and outer spiral arms. At low resolution, the velocity gradients and gas surface densities appear similar in the inner bar and outer spiral arms, perhaps because of beam dilution (e.g. \citealp{kun07}).

We quantitatively study the star formation rate (SFR) and efficiency (SFE = SFR/$M_{\text{H}_2}$; the ratio of the SFR to the molecular gas mass) at a high resolution over the disk of the barred galaxy NGC 4303, using new molecular gas data at high quality. We capitalize on the recent development of the Combined Array for Research in Millimeter-wave Astronomy (CARMA) interferometer, and the 25 beam receiver (BEARS: \citealp{sun00}) and on-the-fly (OTF) mapping technique (\citealp{saw08}) at the Nobeyama 45 m radio telescope (NRO45). The combination of CARMA and NRO45 solves the missing flux problem, providing the best image fidelity and quality, and enabling accurate measurement of gas surface density. 
Our measurement is based on the fundamental transition $J=1-0$ of CO, which has often been used to derive the gas surface density (e.g., \citealp{sco87, you91}). We occasionally refer to ``molecular gas'' as ``gas'' for simplicity in the rest of the paper.

We also investigate the star formation activity in the context of the Kennicutt-Schmidt (KS) law
(\citealp{sch59, ken98}). The KS law is the observational correlation between the area averaged SFR ($\Sigma_{\text{SFR}}$ [$M_{\odot}$ yr$^{-1}$ kpc$^{-2}$]) and gas surface density ($\Sigma_{\text{H}_2}$ [$M_{\odot}$ pc$^{-2}$]) as a power law: $\Sigma_{\text{SFR}} \propto \Sigma_{\text{H}_2}^N$. This correlation of \citet{ken98} spans a wide range, 5 orders of magnitude in $\Sigma_{\text{H}_2}$ and 7 in $\Sigma_{\text{SFR}}$, with $N$ = 1.4 $\pm$ 0.15 using a sample of 61 nearby spiral and 36 starburst galaxies.

Most recent studies concentrate on the KS law at high spatial resolutions, and use local gas and star formation rate densities rather than the averages over galactic disks. The power law index is a topic of interest, since it could indicate the physical origin of the correlation. A linear correlation, $N =1$, would indicate that the star formation efficiency is a constant over a large range of gas surface density; in other words, the mere presence of the gas is sufficient for star formation. On the other hand, a super-linear correlation, $N \neq 1$, may suggest that additional conditions or trigger mechanisms, such as gas dynamics, also play critical roles. 
\citet{won02} averaged their data azimuthally as a function of galactic radius for 7 gas rich galaxies, and obtained $N = 1.1$ and 1.7 (for two different extinction models). 
\citet{ken07} used the locally averaged parameters (520 pc aperture) around H$\alpha$ and 24 $\mu$m peaks in M 51, and obtained $N = 1.56 \pm 0.04$. 
\citet{big08} studied the KS law for 18 nearby galaxies at 750 pc resolution and derived $N =1.0 \pm 0.2$. 

Recently, \citet{ver10} showed that the KS law breaks down at a smaller scale (180 pc).
This may indicate that the mechanism for triggering star formation operates at a scale
smaller than 500 pc, and that star formation is triggered by the environment of GMCs at a scale $\leq$ 500 pc. We study the relation between the gas and star formation at a scale of $\sim$ 250 pc.

\object{NGC 4303} is a nearby barred spiral galaxy with face-on geometry and is the ideal target for this study. It is a member of the Virgo cluster (D = 16.1 Mpc; \citealp{fer96}), whose bar and spiral arms are resolved spatially and kinematically with CARMA (Table \ref{table:4303}). Even a glance at an H$\alpha$ image is enough to recognize the striking difference in star formation activity between the bar and spiral arms in this galaxy (see \citealp{koo01} Figure 5). The CO emission from \object{NGC 4303} has been observed with various radio telescopes, i.e., the Owens Valley Radio Observatory (OVRO: \citealp{shci02}), Nobeyama Millimeter Array (NMA: \citealp{kod06}), Berkeley-Illinois-Maryland Association (BIMA: \citealp{hel03}), and the NRO45 (\citealp{kun07}). We observe \object{NGC 4303} with NRO45 and CARMA and combine the data to obtain a high fidelity image. We focus on discussion of star formation rate (SFR) and efficiency (SFE) as a tracer of star formation activity at high spatial resolution within the galaxy. 

The outline of this paper is as follows. In $\S$2, we present our CO(J=1-0) observations and data analysis. Results are described in $\S$3. SFR and SFE in the bar and spiral arms of NGC 4303 are discussed in $\S$4. The KS law is discussed in $\S$5. A summary is given in $\S$6.

\section{OBSERVATIONS}
We observed the $^{12}$CO(J=1-0) line emission ($f_\text{rest}$ = 115.2712 GHz) toward the barred spiral galaxy NGC 4303 with NRO45 and CARMA. The combination of NRO45 and CARMA interferometer data provides an unprecedented level of image fidelity at a high resolution. Interferometers have an intrinsic limitation: they cannot obtain the shortest spacing components in $u-v$ plane, and thus are not sensitive to the most extended structures. The single-dish NRO45 covers the short spacing data. The combination of the two telescopes is ideal for the study of extended nearby galaxies, enabling accurate measurements of surface brightnesses and gas surface densities. The observing parameters are listed in Table 2.

\subsection{NRO45}
The single dish NRO45 observations were carried out in March 2008 at the Nobeyama Radio Observatory. The twenty five-beam array, BEARS (\citealp{sun00}), consists of double-side band (DSB) superconductor-insulator-superconductor (SIS) receivers on a 5 $\times$ 5 grid with a separation of $\approx$ 41$\farcs$1 at 115 GHz. 
We used the digital auto-correlator for the NRO45 (\citealp{sor00}) with a bandwidth of 512 MHz ($\sim$ 1340 km s$^{-1}$) and 500 kHz resolution (1.3 km s$^{-1}$ at 115 GHz).

The chopper wheel method was used for atmospheric and antenna ohmic loss corrections and to calibrate the antenna temperature in the double-sideband \textit{T}$_\text{A}^{*}$(DSB). A typical system temperature \textit{T}$_\text{sys}$ was 300 -- 400 K in the DSB during our observations. 
\textit{T}$_\text{A}^{*}$(DSB) was converted to the single sideband (SSB) temperature \textit{T}$_\text{A}^{*}$(SSB) by applying scale factors (hereafter, we use \textit{T}$_\text{A}^{*}$ to refer to \textit{T}$_\text{A}^{*}$(SSB)). The scale factors were measured by comparing images of the Orion IRc2 region in the CO(J=1-0) line obtained using BEARS and the single-beam SSB receiver S100. 
The main beam efficiency of NRO45 was $\eta_{\text{MB}}$ = 0.30 at 115 GHz. The main beam temperature is calculated as \textit{T}$_{\text{MB}}$ = \textit{T}$_\text{A}^{*}$/$\eta_{\text{MB}}$. 

\begin{figure}[h]
%\epsscale{.80}
\includegraphics[width=10cm]{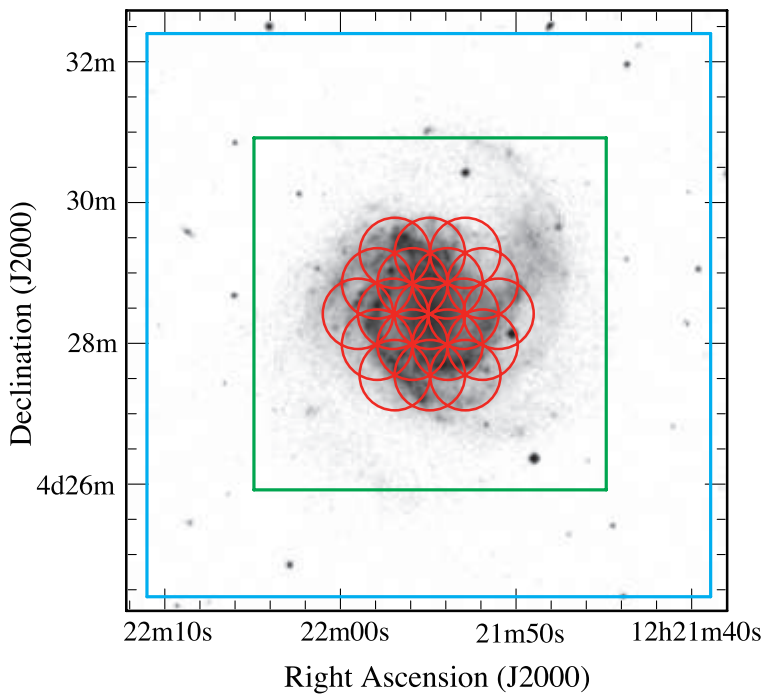}
\caption{Optical image of NGC 4303 overlaid with regions observed using the NRO45 (blue outline) and CARMA (red circles). The 25 BEARS beams are spread evenly inside the region outlined in green, which we used for data reduction.}
\label{fig:OBSERVATION}
\end{figure}

The telescope pointing was checked roughly every 30 minutes by observing the nearby SiO maser source R-Crv with another SIS receiver, S40 near 40 GHz. 
The standard deviation of the pointing offset between the optical axes of S40 and BEARS was 0$\farcs$87, which is measured by the observatory. 
The pointing error was less than 6$\arcsec$ during our observations. 
We mapped the 8$\arcmin$ $\times$ 8$\arcmin$ region centered on \object{NGC 4303} with the OTF observing mode (\citealp{saw08}), which corresponds to a 37 kpc $\times$ 37 kpc area at the distance of \object{NGC 4303}. 
This area covers the optical extent of the galaxy. We alternated between continuous scans in RA and DEC, as systematic noise appears dominantly along the direction perpendicular to the scan, which causes it to cancel out when the scans are combined.
The 25 beams draw a regular 5$\arcsec$ stripe in a single scan, which is finer than the Nyquist spacing of the 15$\arcsec$ NRO45 beam --- no spatial information is missed, which is critical in producing good $u-v$ coverage and in combining with CARMA data. 
We construct data cubes only for the 5$\arcmin$ $\times$ 5$\arcmin$ area covered by all 25 beams (Figure \ref{fig:OBSERVATION}).

We used the data reduction package \textit{NOSTAR} (Nobeyama OTF Software Tools for Analysis and Reduction) developed at NRO (\citealp{saw08}). 
After the DSB-to-SSB conversion, we subtracted spectral baselines with a first order polynomial fit and flagged bad integrations. 
The RA and DEC scans provide data points roughly along a 5$\arcsec$ lattice. We used a spheroidal function to re-grid them onto a regular grid to produce an integrated data cube. 
The \textit{Spheroidal function} has low sidelobes in the Fourier domain and is advantageous in de-convolving the map to produce $u-v$ data. 
The convolution smears out the resolution of the map produced by the 15$\arcsec$ NRO45 beam to 22$\farcs$1. 
We produced separate data cubes for the scans taken in the RA and DEC directions and combine them using the basket-weave method in \textit{NOSTAR}. 
Lower weights were applied to the data at small wave-numbers in the scan directions in the Fourier domain, which significantly reduces the systematic noise in the combined map. 

The final data cube has a spatial resolution of 22$\farcs$1 and a pixel scale of 7$\farcs$5, which is the half size of 15$\arcsec$ NRO45 beam. The rms noise is 75 mK at the velocity resolution of $\Delta V$ = 2.6 km s$^{-1}$.

\subsection{CARMA}
We observed the central 2$\farcm$3 region of \object{NGC 4303} with CARMA in the C-configuration (October and November 2007) and in the D-configuration (June 2008). 
CARMA is a 15-element array, which consists of nine 6.1-m antennas (formerly BIMA) and six 10.4-m antennas (formerly OVRO). 
The angular resolution and baseline coverage of C-configuration (D-configuration) is 2$\arcsec$ (5$\arcsec$) and 26 -- 370 m (11 -- 148 m). 
The spatial resolution corresponding to the minimum baseline (11 m) is 49$\arcsec$. 
We adopted a 19 point hexagonal mosaic pattern with 30$\arcsec$ spacing (i.e., the Nyquist spacing of a 10.4-m primary beam; Figure \ref{fig:OBSERVATION}). We used a 3 mm SIS receivers, and digital correlators with 3 bands, each of which has 62 MHz of bandwidth in a single side band with 977 kHz resolution (2.5 km s$^{-1}$ at 115 GHz). 

The bright quasar, 3C273, was observed for bandpass and gain calibrations. We observed it roughly every 20 minutes to calibrate for atmospheric and instrumental gain variations. 
We corrected the telescope pointing by observing 3C273 with the 3 mm receiver. 
We frequently monitored and corrected the pointing during tracks, using the optical pointing method developed by \citet{cor08}, which involves observing bright optical stars close to 3C273 every gain calibration cycle (around twenty minutes). 
The pointing offset vector between the radio and optical pointings was measured at the beginning of the observations. 

We used the data reduction package \textit{MIRIAD} (\citealp{sau95}) for calibration and data analysis. The rms noise level of the CARMA data is 43 mJy beam$^{-1}$ in a channel of $\Delta V$ = 2.6 km s$^{-1}$. The synthesized beam size is 3$\farcs$1 $\times$ 2$\farcs$5.

\subsection{Combining NRO45 and CARMA data in $u-v$ space}
We combined CARMA and NRO45 data in the Fourier domain as done in \citet{kod09, kur09}. The NRO45 provided the short spacing data which filled the central hole in the CARMA $u-v$ coverage. The combined image achieved higher angular resolution than the NRO45 map and recovered the flux missing in the CARMA map. 

\begin{figure}[h]
\begin{center}
\includegraphics[width=10cm]{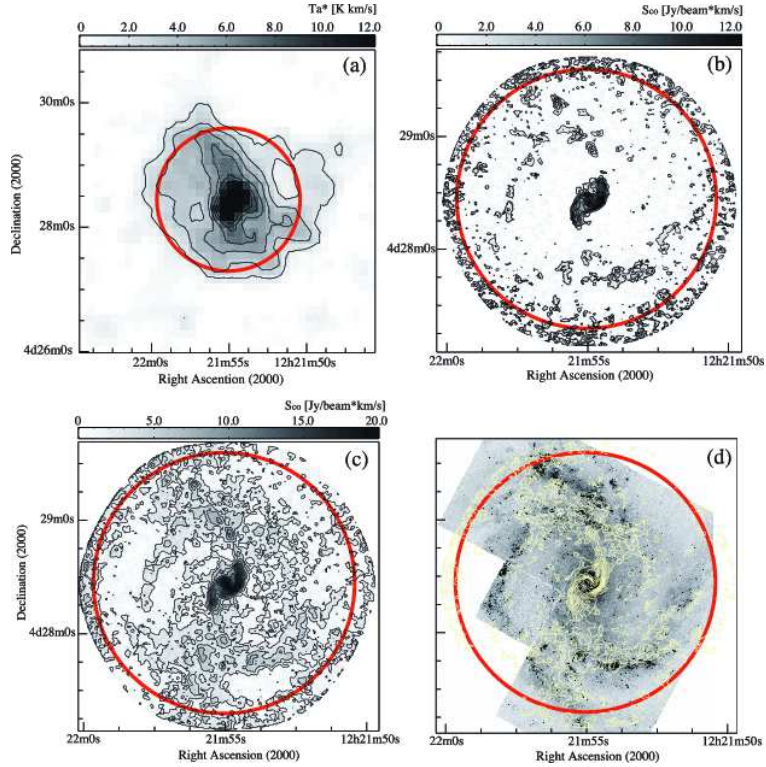}
\caption{Integrated intensity map for (a) NRO45, (b) CARMA, (c) NRO45 + CARMA (combined data), and (d) combined data overlaid on HST F450 image. The contours trace 10$\%$ intervals from 10$\%$ to 90$\%$ of the peak. 1$\sigma$ noise levels are 1.5 K km s$^{-1}$ (NRO45), 810 mJy beam$^{-1}$ km s$^{-1}$ (CARMA), and 710 mJy beam$^{-1}$ km s$^{-1}$ (combined data). The peak intensities of the NRO45, CARMA and NRO45 + CARMA images are 15.7 K km s$^{-1}$, 13.0 Jy beam$^{-1}$ km s$^{-1}$, and 20.8 Jy beam$^{-1}$ km s$^{-1}$, respectively. The NRO45 + CARMA image achieves a high resolution as well as recovering extended emission components. The CO emission overlaps with dust lanes, which are seen as extinction in the HST image. Red circles are the central 2$\farcm$3 region observed with CARMA.}
\label{fig:MOM0}
\end{center}
\end{figure}

By utilizing \textit{MIRIAD\footnote{see http://bima.astro.umd.edu/miriad/}} and new routines that we developed, we combined NRO45 and CARMA data in the following way: \\ 
\begin{enumerate}
\item {Deconvolution of single dish map with single-dish beam} \\
The point spread function (PSF) of the NRO45 map is a convolution of the NRO45 beam pattern and the spheroidal function that we applied to regrid the data points and create the final data cube. The map should be deconvolved by the PSF before being Fourier-transformed to the $u-v$ plane. 
The NRO45 beam size is Gaussian with an effective beam size of 22$\farcs$1.
\item {Multiplying dummy 2$\arcmin$ Gaussian primary beam} \\
This step produced the same data structure for the single dish data as is used by the interferometer data in order for the files to be read into \textit{MIRIAD}. 
We used a 19 point hexagonal mosaic technique in our CARMA observations, so the NRO45 map was divided into 19 smaller maps corresponding to CARMA's mosaic points. 
The pseudo-primary beams were multiplied by each of the 19 maps to mimic CARMA observations. 
CARMA has three types of primary beam pattern with a FWHM of 64$\arcsec$ for 10.4 m -- 10.4 m, 85$\arcsec$ for 6.1 m -- 10.1 m, 115$\arcsec$ for 6.1m -- 6.1 m at 115 GHz. 
To mimic this behavior, we applied a Gaussian primary beam with a FWHM of 2$\arcmin$ to the deconvolved NRO45 map at each of the 19 pointings and produced 19 maps. The Fourier transformation of these maps are the $u-v$ data we need (see step 4). 
\item {Determining weightings} \\
The relative weighting of NRO45 and CARMA data changes the synthesized beam pattern. By changing the weighting, we can control the sidelobes of the synthesized beam and the noise level. This is similar to the conventional weighting schemes (e.g. natural and uniform weighting, or robust parameter). We decided to use the weighting that minimized the sidelobes. 
\item {Constructing visibility data from NRO45 data} \\
All of the visibility data of NRO45 is built after applying the relative weight balance between the visibility data of NRO45 and CARMA \citep{kur09}. We made 57 visibility data sets, i.e., for 19 pointings $\times$ 3 primary beams. All 57 NRO45 visibility data sets are concatenated to produce the final NRO45 $u-v$ data. 
\item {Imaging} \\
We obtain the combined data cube by adding together the visibility data sets from NRO45 and CARMA, and by Fourier transforming the result, we created a combined image. 
The total flux of the combined image is similar to that of the NRO45 image. 
The rms noise of the combined cube is 34 mJy beam$^{-1}$ with $\Delta V$ = 2.6 km s$^{-1}$. The synthesized beam is 3$\farcs$2 $\times$ 3$\farcs$2.
\end{enumerate}
Bright CO emission is detected throughout the disk of \object{NGC 4303}, even in the inter-arm region in NRO45 map (Figure \ref{fig:MOM0} (a)). On the other hand, the CARMA map shows emission only from a compact component in the nuclear region and the spiral arms (Figure \ref{fig:MOM0} (b)). The combined data map clearly shows strong CO emission in the nucleus, the bar, and the disk region along the two arms (Figure \ref{fig:MOM0} (c)). A comparison of the total flux in the  NRO45, CARMA, and combined images clearly shows that the missing flux has been completely recovered (Table \ref{table:comparison}).

\subsection{Archival data}
We obtained archival H$\alpha$ (\citealp{kna04}) and K-band \citep{mol01} images from the NASA Extragalactic Database (NED\footnote{see http://nedwww.ipac.caltech.edu/}) and a 24 $\mu$m image (P.I. Kotak) from the Spitzer Space Telescope data archive using the Leopard software. The retrieved 24 $\mu$m image suffered from residual sky background emission, which we subtracted using the Image Reduction and Analysis Facility (\textit{IRAF\footnote{IRAF is distributed by the National Optical Astronomy Observatories, which are operated by the Association of Universities for Research in Astronomy, Inc., under cooperative agreement with the National Science Foundation.}}) package.

UV photons from recently-formed, young, massive stars primarily ionize the surrounding gas (\ion{H}{2} regions) or heat dust grains (thermal radiation). Thus, the combination of H$\alpha$ and 24 $\mu$m dust continuum emissions traces the star formation rate fairly accurately \citep{cal07, ken07}.
The SFR is calculated using the formulae from \citet{ken07} and \citet{cal07}, namely
\begin{equation}
\begin{split}
&\text{SFR}  \text{[M$_{\sun}$ year$^{-1}$]} = 7.9 \times 10^{-42} L(\text{H$\alpha$})_{corr}  \text{[ergs s$^{-1}$]} \\
					 &= 7.9 \times 10^{-42} ( L(\text{H$\alpha$})_{obs} + (0.031 \pm 0.006) L(24 \mu m) )  \text{[erg s$^{-1}$]},
					 \label{eq:SFR}
\end{split}
\end{equation}
where $L$(H$\alpha$) and $L$(24 $\mu$m) are the H$\alpha$ and 24 $\mu$m luminosities.
From the H$\alpha$ and 24 $\mu$m images, we found that H$\alpha$ emission defined the sites of star forming regions very well, although it does not quantitatively trace the star formation rate (SFR). Since the H$\alpha$ image has a much higher resolution ($1\farcs5$) than the 24 $\mu$m image, we use the H$\alpha$ image in discussing the offsets between star forming regions and parental gas qualitatively. For quantitative measurements, we convolve the H$\alpha$ data, and smooth the H$\alpha$ and CO images to match the spatial resolution of the 24 $\mu$m image (6$\arcsec$ $\sim$ 500 pc). 
We clip the SFR and CO data at the 2$\sigma$ r.m.s. noise level. The SFR map is shown in Figure \ref{fig:SFR_SFE} (left).

\section{Results}
\subsection{Molecular Gas Distribution}
Molecular gas distributions in barred spiral galaxies have been studied by many authors, from high resolution observations of central bars (e.g. \citealp{ken93, ish99, sak99a, sak99b}) to lower resolution observations of larger disks \citep{she00}. 
The gas distribution in \object{NGC 4303} is typical among barred spiral galaxies; it shows two narrow ridges (namely, offset ridges) at the leading side of the bar, and spiral arms in the outer part, and coincides with dust lanes in optical images. 
Figure \ref{fig:MOM0} (d) shows an overlay of CO contours on an archival HST F450 band image (P.I. Smartt).

The CO emission is concentrated in the central 40$\arcsec$ ($\sim$ 3.2 kpc), which includes the offset ridges and circumnuclear disk.
In particular, the circumnuclear disk shows a circular structure ($r$ = 4$\farcs$4 $\sim$ 350 pc), extending from 1541.3 km s$^{-1}$ southeast of the nucleus to 1611.5 km s$^{-1}$ to the northwest in the velocity channel maps (Figure \ref{fig:CH_COMB1}, Figure \ref{fig:CH_COMB2}). 
This structure is called a ring or part of the offset ridges  \citep[i.e., inner spiral arms; ][]{shci02, kod06}. The circumnuclear disk hosts a spiral arm in the UV continuum image  (\citealp{col97}), and the CO ring/spiral arm structure is offset from the UV spiral arm.

Low-level CO emission is also present in the space outside of the offset ridges (namely, inter-offset ridge regions). 
Strong concentrations of gas exist at the outer ends of the offset ridges (Figure \ref{fig:MOM0} (c)). Molecular spiral arms extend from those end-points toward the outer regions. The map resolves the widths of the offset ridges of the bar and the outer spiral arms at high resolution. 
The surface densities of the gas in the offset ridges and outer spiral arms are not similar at the high resolution, differing by factor of $\sim$ 2.

We derive the inclination and position angle of the galaxy using the task $GAL$ in the AIPS data reduction package. The results are shown in Table 1. We use these values for the rest of our analysis.

\subsection{Molecular Gas Mass and Surface Density}
We estimate the gas mass from the CO integrated intensity $S_{\text{CO}}$ (see Figure \ref{fig:MOM0} (c)), using the Galactic \textit{I}(CO)-to-\textit{N}(H$_2$) conversion factor \textit{X}$_\text{CO}$. We adopt the standard value $X_{\text{CO}}$ = 2.0 $\times$ 10$^{20}$ cm$^{-2}$ [K km s$^{-1}$]$^{-1}$ (\citealp{str87}). The gas mass is then
\begin{equation}
\begin{split}
M_{\text{H}_2} = &1.0 \times 10^4 
\left( \frac{S_\text{CO}}{\text{Jy km s$^{-1}$}} \right) \left( \frac{D}{\text{Mpc}} \right)^2 \\
&\left[ \frac{X_\text{CO}}{3.0 \times 10^{20} \text{cm$^{-2}$ (K km s$^{-1}$)$^{-1}$}} \right] M_{\sun},
\end{split}
\end{equation}
where \textit{D} is the distance to the galaxy , which assumed to be 16.1 Mpc (\citealp{fer96}).
The total gas mass of the galaxy is calculated from the NRO45 map to be $5.3\times 10^9$ M$_{\sun}$. 
The CARMA mosaic covered the circular area with a radius of $80\arcsec$. The enclosed mass is $3.4\times 10^9$ M$_{\sun}$. 
The average gas surface density in the central $160\arcsec$ of the molecular disk is 36 M$_{\sun}$ pc$^{-2}$.

\section{Molecular Gas and Star Formation}
\subsection{Spatially Resolved Sites of Star Formation} \label{sec:resolvedSF}
Both CO and H$\alpha$ emission appear clearly along stellar spiral arms, and the H$\alpha$ arms are offset toward the leading side of the CO arms. Similar offsets are evident between CO and 24 $\mu$m emission. The 24 $\mu$m emission is at the leading side of the CO emission. These offsets are due to the delay of star formation from the accumulation of gas in the spiral arms (\citealp{egu04}); approximately, an offset would be $(\Omega-\Omega_\text{p}) t_{\it SF}$, where $\Omega$ and $\Omega_\text{p}$ are the angular speed of the gas and pattern speed of spiral arms, respectively, and $t_{\it SF}$ is the timescale of star formation. 
The sites of star formation are resolved in our observations. 
The analysis of local star formation rate and efficiency should take this into account, since the parental gas and star forming regions would not be correlated at the high spatial resolution. 
This means that the KS law breaks down at this resolution. We need to smooth the data before characterizing the spatial variation of the star formation efficiency in \S 4.2.

Figure \ref{fig:off} (a), (b) show a close-up of the offset between CO (black contours) and H$\alpha$ emissions (greyscale). In the upper panels, the arrows indicate the direction of gas flow, assuming pure azimuthal rotation and a time lag between star formation and gas accumulation in spiral arms. Figure \ref{fig:off} (c) shows the phase diagrams of CO (contours) and H$\alpha$ (greyscale). Again, the H$\alpha$ emission is found predominantly at the leading side of the CO spiral arms over the entire range of radii (over 40$\arcsec$, $\sim$ 3.1 kpc). 
 
The offsets between the CO and H$\alpha$ arms  range from 0 to 1 kpc, with an average of $\sim 500$ pc at  radius, $r$ $\geq$ 40$\arcsec$ ($\sim$ 3.1 kpc; i.e., the area we define as outer spiral arms in \S 4.4). Thus, we will smooth the data to the spatial resolution of $\sim$ 500 pc for later discussions; it also especially affects the analysis of SFE. We will discuss this further in $\S$ 4.4.

\citet{egu04} introduced a new method to estimate the timescale of star formation using these offsets, measuring the offset/time lag between gas accumulation and star formation. 
\citet{tam08} applied this method to \ion{H}{1} and 24 $\mu$m data. Their offsets $\leq$ 250 pc (calculated from their paper), are translated to the timescale of $\sim$ 1 Myr. 
On the other hand, \citet{egu09} used CO and H$\alpha$ data for NGC 4303, and derived the offsets of 700--1000 pc and the timescale of $\sim$ 10 Myr, an order of magnitude longer than the \citet{tam08} results. Our offsets are closer to those in \citet{egu09}. 
To understand this discrepancy, we compared 24 $\mu$m and H$\alpha$ images, and found no significant offsets between the two. Thus, H$\alpha$ traces the locations of star forming regions as well as 24 $\mu$m. 
The discrepancy may come from the difference of CO and \ion{H}{1}. If the \ion{H}{1} emission traces the gas dissociated by star formation, and not the natal gas for star formation, it explains the small offsets between \ion{H}{1} and 24 $\mu$m emissions. Our simple analysis with CO and H$\alpha$/24 $\mu$m favors the longer star formation timescale of \citet{egu09} in case of NGC 4303. These results should be confirmed with a larger sample of galaxies.

\subsection{Star Formation Rate and Efficiency}
We compare the star formation rate (SFR) and star formation efficiency (SFE) between the bar and outer spiral arms, as well as within other components (i.e., circumnuclear disk, inter-arm regions), at the spatial resolution of $\sim 500$ pc.
The SFE is the ratio of the SFR to the gas mass within an aperture, namely
\begin{equation}
\text{SFE} [\text{yr$^{-1}$}] = \frac{\text{SFR} [\text{M}_{\sun} \text{yr$^{-1}$}]}{M_{\text{H}_2} [\text{M}_{\sun}]} , \label{eq:SFE}
\end{equation}
where $M_{\text{H}_2}$ is the molecular hydrogen gas mass. 
We calculate the SFR and gas mass within an aperture and use the averaged parameters to derive the SFE. 
We smooth our CO and H$\alpha$ images to the same resolution as the 24 $\mu$m image (6$\arcsec$ resolution) in order to correlate the parental gas and star forming regions. This resolution corresponds to a linear size of $\sim$ 500 pc, similar to the linear resolution at which \citet{ken07} find the correlation between gas surface density and associated star formation rate. 

\begin{figure}[h]
\includegraphics[width=10cm]{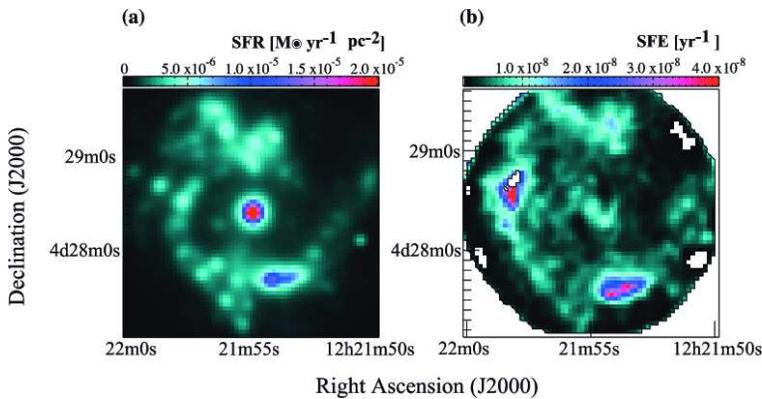}
\caption{SFR map (left) and SFE map (right) of NGC 4303. SFR is calculated from the H$\alpha$ and Spitzer 24 $\mu$m images so that the effect of extinction is small. }
\label{fig:SFR_SFE}
\end{figure}

\begin{figure}[h]
\includegraphics[width=15cm]{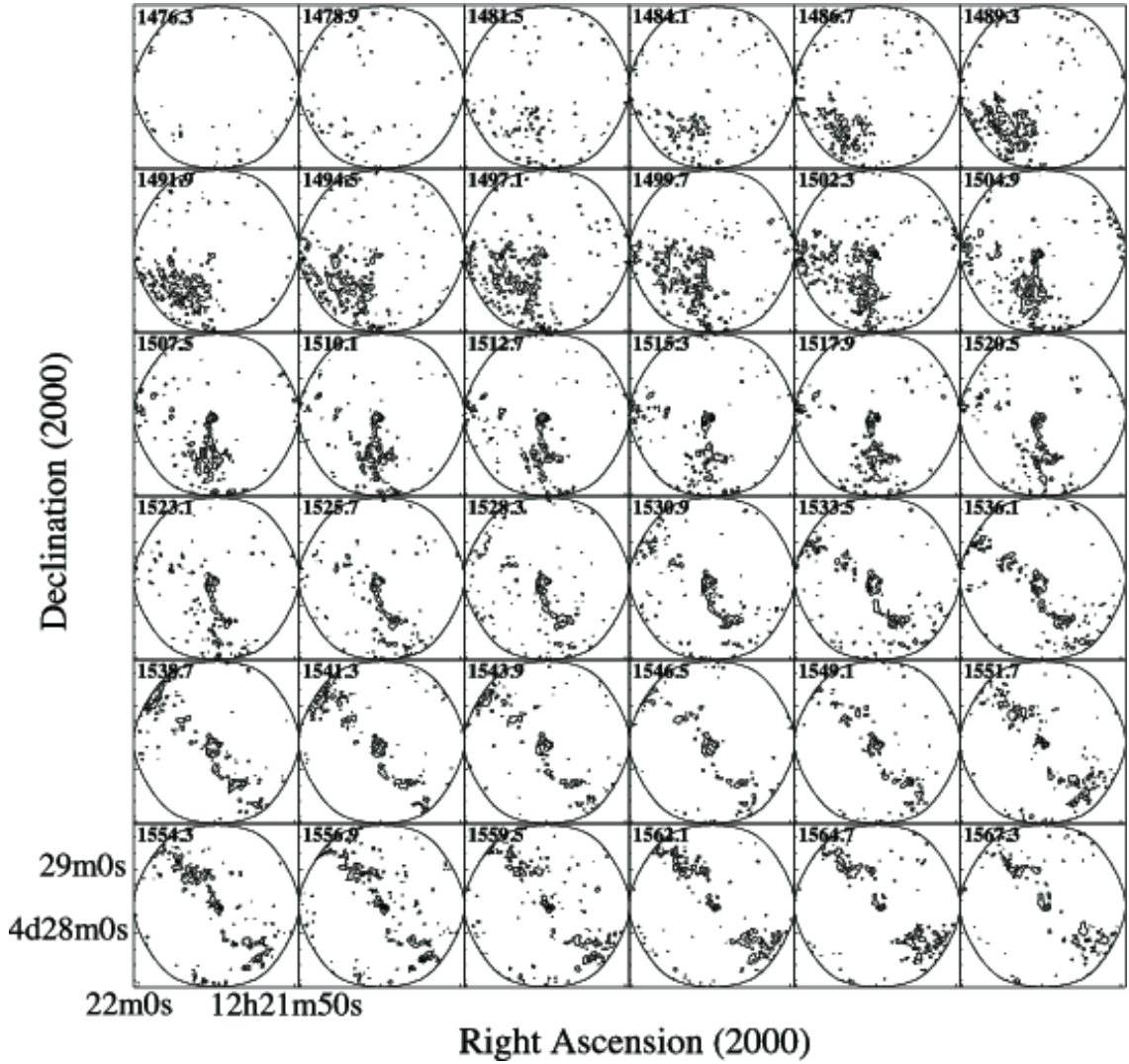}
\caption{Channel maps of the combined data from 1475.0 km s$^{-1}$ to 1566.0 km s$^{-1}$. The contours are 3, 6, 9$\sigma$, where the 1$\sigma$ noise level is 34 mJy beam$^{-1}$.}
\label{fig:CH_COMB1}
\end{figure}

\begin{figure}[h]
\includegraphics[width=15cm]{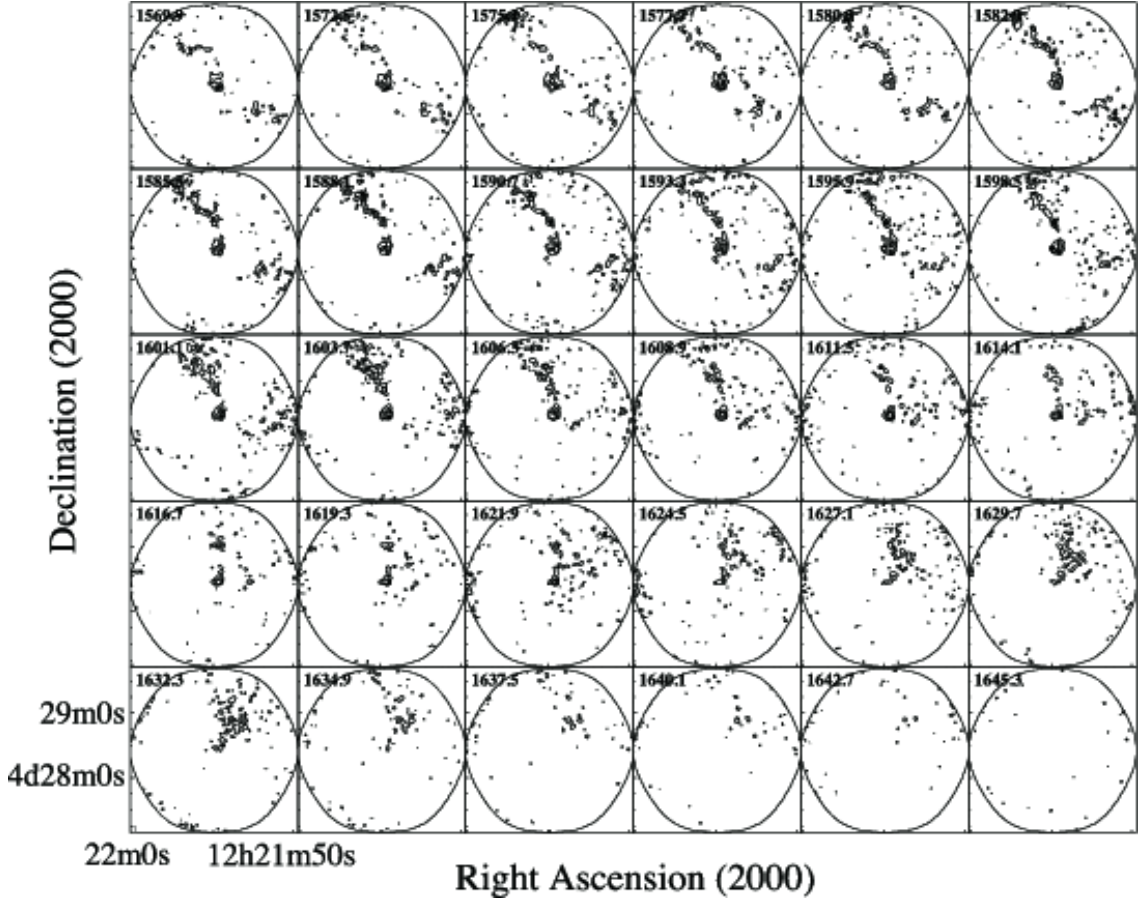}
\caption{Channel maps from 1568.6 km s$^{-1}$ to 1644.0 km s$^{-1}$.}
\label{fig:CH_COMB2}
\end{figure}

\begin{figure}[h]
\begin{center}
\includegraphics[width=10cm]{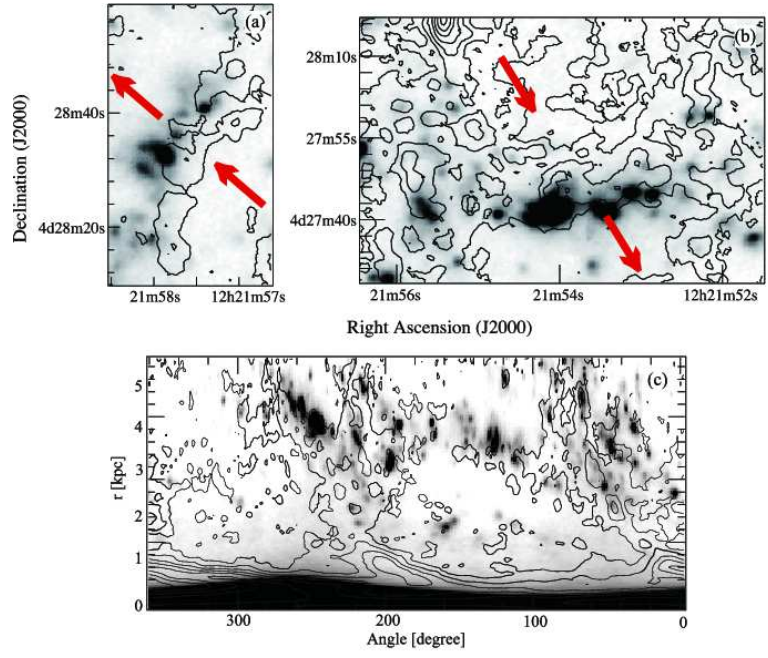}
\caption{(a), (b) The distributions of CO (black contours) and H$\alpha$ (greyscale) emission in two regions of NGC 4303 are shown. The direction of gas flow is determined assuming a circular galactic rotation and that the young massive stars (traced by H$\alpha$) flow downstream from the molecular spiral arms where they were formed. (c) The phase diagram of CO (contours) and H$\alpha$ (grey scale).}
\label{fig:off}
\end{center}
\end{figure}

A SFR map is shown in Figure \ref{fig:SFR_SFE} (a). 
High SFRs are apparent in intense star forming regions which are sparsely distributed along the spiral arms, as well as in the circumnuclear disk. 
However, SFRs  are lower in the bar, even though the gas surface densities along the bar offset ridges are as high as those in the spiral arms. 
The low SFR in the bar has been suggested qualitatively by previous studies (most systematically by \citealp{she02}), and we discuss this quantitatively in the following sections.

Figure \ref{fig:SFR_SFE} (b) shows a map of the SFE. Low SFEs are seen in the bar (especially around the offset ridges) and in the inter-arm regions. Exceptionally high SFEs appear at the leading side of the CO arms. This indicates that the offsets between H$\alpha$ and CO emission are not fully smoothed out even at the resolution of $\sim 500$ pc (see $\S$ 3.2), even though the parental gas shows a fair correlation with SFR \citep{ken07}.

\subsection{Azimuthally Averaged Star Formation Efficiency}
\begin{figure}[h]
\begin{center}
\includegraphics[width=7cm]{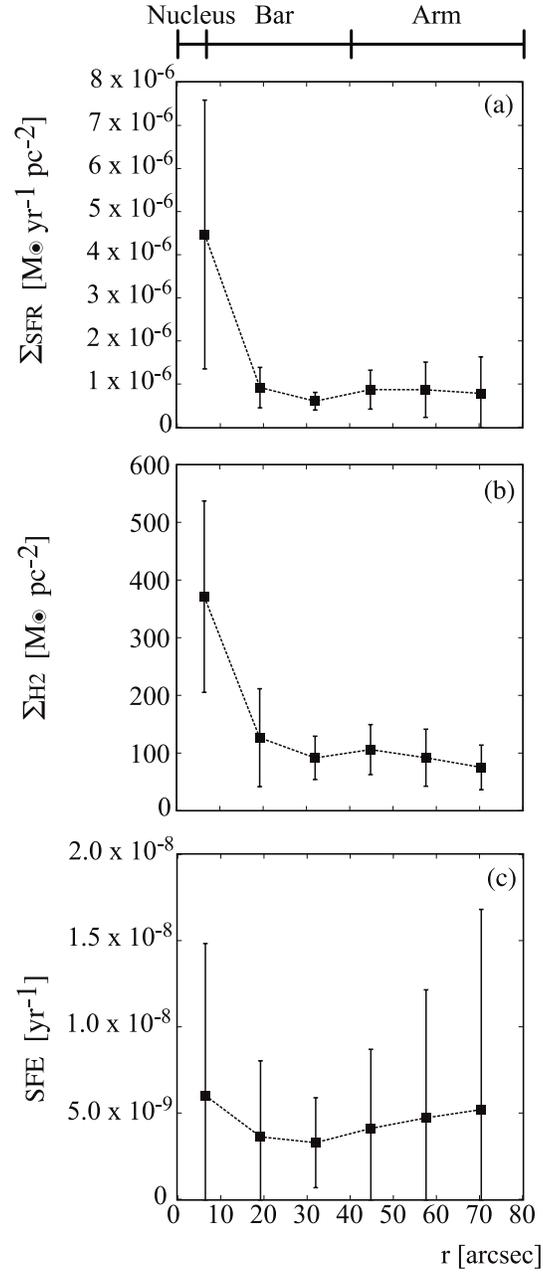}
\end{center}
\caption{Radial distributions of (a) $\Sigma_{\text{SFR}}$;  (b) $\Sigma_{\text{H}_2}$; (c) SFE. The error bars are one r.m.s.}
\label{fig:radial}
\end{figure}

We first compare the gas surface density ($\Sigma_{\text{H}_2}$), area averaged SFR ($\Sigma_{\text{SFR}}$), and SFE between the bar and spiral arms on the basis of azimuthal averages (Figure \ref{fig:radial}). The average values are derived in 1 kpc (12$\farcs$8) width annuli, which is enough to average out the spatial offsets (see $\S$ 3.2). The 1kpc resolution can still separate the bar, spiral arms, and nucleus without significant contamination from each other. Figure \ref{fig:point} (a) shows the definitions of nucleus, bar, and outer spiral arms.
The nucleus might include a portion of the bar region; however, we are mainly interested in the differences between the bar and spiral arms, so it does not affect our analysis. We will comment on the properties of the circumnuclear star formation in $\S$ 4.4.

Figure \ref{fig:radial} shows the radial distributions of the area averaged SFR $\Sigma_{\rm SFR}$ (a), gas surface density $\Sigma_{\rm H_2}$ (b), and SFE (c). 
The definitions of the nucleus, bar, and spiral arms are indicated with arrows in Figure \ref{fig:radial} (a). SFE drops abruptly from the circumnuclear disk to the bar, and increases toward the spiral arms (Figure \ref{fig:radial} (c)). Quantitatively, the average SFE is about twice as high in the spiral arms as in the bar.

\subsection{Locally Averaged Star Formation Efficiency}
The star formation efficiency (SFE, Eq. \ref{eq:SFE}) is a correlation between star forming regions and their parental gas. We have resolved the scale where the star forming regions and gas are spatially separated; therefore, we have to smooth the data to derive a SFE with physical importance.  \citet{ken07} found a good correlation between the surface densities of SFR and gas mass at the 500 pc scale. 
This is similar to the typical separation of H$\alpha$ and CO emission that we see in NGC 4303 (\S \ref{sec:resolvedSF}). We smooth H$\alpha$ and our CO images to the scale of the Spitzer 24 $\mu$m image ($6\arcsec$; $\sim 500$ pc), and compare local SFR and SFE between the bar and spiral arms at this scale.

\begin{figure}[h]
\begin{center}
\includegraphics[width=10cm]{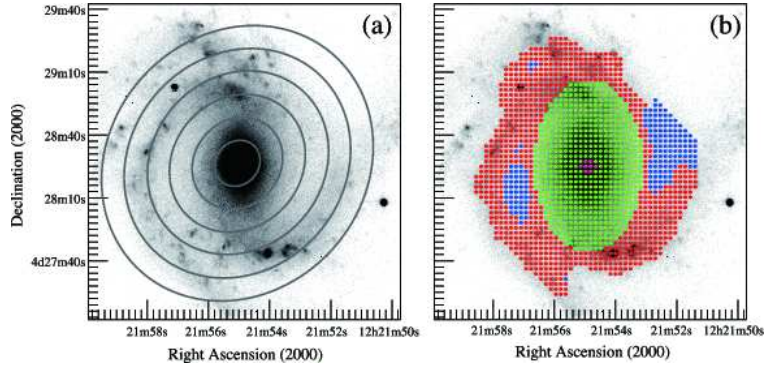}
\end{center}
\caption{Definitions of the nucleus, bar, and spiral arm (outer disk) regions. Left: azimuthal definitions. Each annulus has 1 kpc width. Right: local definitions with 3$\arcsec$ grid. The nucleus, bar, arm and inter-arm are represented by pink, green, red and blue, respectively.}
\label{fig:point}
\end{figure}

Four regions (i.e., bar, spiral arms, inter-arms regions, and circumnuclear region) are defined as in Figure \ref{fig:point} (b). The circumnuclear region is a 6$\arcsec$ diameter circle ($\sim 500$ pc), based on a near-infrared (NIR) image of P\'{e}rez-Ram\'{i}rez et al. (2000). We define the bar as an ellipse with the deprojected major and minor axis diameters of 40$\arcsec$ $\times$ 26$\arcsec$ (3.1 $\times$ 2.0 kpc), based on a K-band image. This definition yields an ellipticity similar to the one derived by \citet{lai02} based on a near infrared $H$-band image, and a major axis diameter consistent with the determination of  \citet{shci02}. The spiral arms and inter-arm regions are separated by eye using the K-band image.  Figure \ref{fig:point} (b) shows our definitions overlaid on the $K$-band image.

\begin{figure}[h]
\begin{center}
\includegraphics[width=16cm]{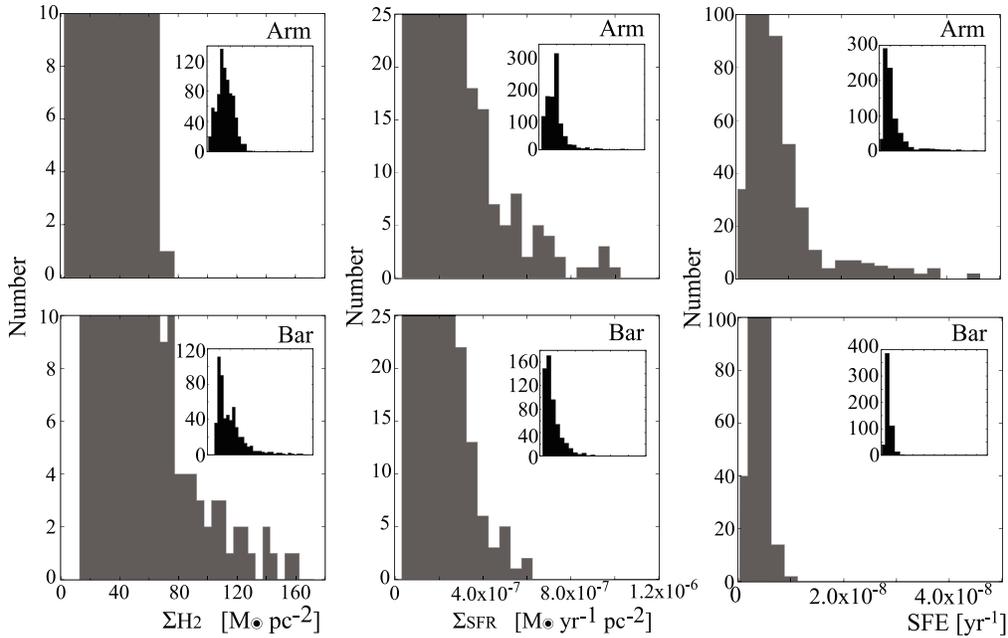}
\end{center}
\caption{Histograms of (left) surface density $\Sigma_{\text{H}_2}$, (middle) area averaged star formation rate $\Sigma_{\text{SFR}}$, and (right) SFE. The insets are overall view of the histograms.}
\label{fig:hist}
\end{figure}

Figure \ref{fig:hist} shows the histograms of $\Sigma_{\text{H}_2}$ (left), $\Sigma_{\text{SFR}}$ (center), and SFE (right) for the spiral arms and bar (top and bottom). High gas surface density regions are present only in the bar, but not in the spiral arms. However, high SFR and SFE regions appear only in the spiral arms and are absent in the bar. Therefore, high SFR and SFE are evident in the spiral arms in the local average, as well as in the azimuthal average.

We calculate the excesses of SFR and SFE in each region with respect to the averages over the entire disk excluding the nucleus (Table 4). The SFR and SFE in the bar are 10 $\%$ and 30 $\%$ smaller than the disk averages, respectively. Star formation activity in the bar is low compared to the entire disk. On the contrary, the SFR and SFE in the spiral arms are about 10 $\%$ and 40 $\%$ higher than those over the disk, respectively. Active star formation preferentially occurs in the spiral arms.
SFR and SFE are about 30 $\%$ and twice as high in the spiral arms as in the bar. The results confirm quantitatively the notion that star formation in bar is less than in spiral arms.

The effect of the spatial offsets between H$\alpha$ and CO emissions is significantly reduced by the smoothing to estimate local SFE ($\sim$ 500 pc resolution). However, it still shows some residual effects in SFEs, since the average of this offset is also $\sim$ 500 pc. If the offset is below 500 pc, parent GMCs and star forming regions exist within our resolution for SFE discussions. If the offset is above 500 pc, we cannot recognize both parent GMCs and star forming regions within the resolution. SFEs obtained in such a region will not fairly represent the local SFE. This produces SFEs above 10$^{-7.8}$ yr$^{-1}$ are the artifacts (Figure \ref{fig:SFR_SFE}).

\subsection{Circumnuclear Star formation}
The SFR and SFE are the  highest in the nuclear region. The azimuthal averages of $\Sigma_\text{SFR}$ and SFE (Figure \ref{fig:radial} (a), (c)) clearly show this tendency, being about 5 times and twice higher than the disk averages, respectively. The locally averaged SFE in the nucleus is also high. The locally averaged SFR and SFE are higher by a factor of 6.9, and 1.3, respectively than those of the entire disk region.

We, however, note that NGC 4303 is suggested to harbor an AGN in addition to a starburst (e.g. \citealp{ken89, col97,tsc00}), and our analysis could suffer from AGN contamination. 
We checked the spectral energy distribution (SED) of the nucleus in 1--24 $\mu$m, using the archival data (\textit{Catalog of Infrared Observations\footnote{http://ircatalog.gsfc.nasa.gov/}}; \citealp{gez93}). We found a flat spectrum from 10--24 $\mu$m, similar to AGN spectra (\citet{her03}). Thus, 24 $\mu$m emission from the nucleus is dominated by the AGN. This indicates that the much higher values of SFR in the nucleus are overestimated.

\subsection{Comments on metallicity-dependent $X_{\text{CO}}$}
We adopt the standard conversion factor $X_{\text{CO}}$. However, there are some arguments supporting a metallicity-dependent $X_{\text{CO}}$ (e.g. \citealp{ari96, bos02}). \object{NGC 4303} has a relatively large radial metallicity gradient (\citealp{ski96}), which may affect the result that SFE is low in the bar (inner disk) and high in spiral arms (outer disk). We confirmed as follows that this does not change the results (except for the nucleus, where metallicity is exceptionally high).

The average metallicities (12 + $\log$[O/H]) in the nucleus, bar, and spiral arms (the definition is same as Figure \ref{fig:point} (b)) are 9.5, 9.3, and 9.2 from \citet{ski96}. If we adopt the $X_{\text{CO}}$-metallicity relation of \citet{bos02}, the gas surface densities decrease, and the average SFEs increase (by a factor of 4.2, 2.3, and 1.8) respectively. The difference in SFE between the bar and spiral arms is reduced only by 30 $\%$. This is too small to change the trend that the SFE is roughly a factor of two higher in spiral arms than in bar.
A significant effect appears only in the nucleus, where a much higher metallicity results in a significantly higher SFE (a factor of 2 -- 4 higher than that from the standard $X_{\text{CO}}$).

\section{The Kennicutt-Schmidt law}
The Kennicut-Schmidt (KS) law shows a tight correlation between the gas surface density $\Sigma_{\text{gas}}$ and SFR density $\Sigma_{\text{SFR}}$ \citep{ken98} as we mentioned in $\S$ 1. The difference in SFE between the bar and spiral arms may affect this correlation. Figure \ref{fig:ks} (a) shows the surface densities of gas ($\Sigma_{\text{H}_{2}}$ [M$_{\odot}$ pc$^{-2}$]) and area averaged star formation rate ($\Sigma_{\text{SFR}}$ [M$_{\odot}$ yr$^{-1}$ pc$^{-2}$]) for our data at $\sim$ 500 pc resolution. We use the same axis ranges as those in Figure 9 of \citet{ken98}, spanning five orders of magnitude in $\Sigma_{\text{H}_{2}}$ and seven orders of magnitude in $\Sigma_\text{SFR}$, a range which covers from normal spiral galaxies to circumstellar starburst. 

The central region of our data could be contaminated with AGN, and the SFR calculated form 24 $\mu$m could be overestimated.Therefore, we fit straight lines to two sets of data, i.e., with and without the central region, to see the effect of AGN on the KS fits. 
Figure \ref{fig:ks} (a) shows our least square fit results: the orange and blue solid lines are for the data with and without the central region, and are $\Sigma_{\text{SFR}} = 10^{-2.61 \pm 0.04} \times \Sigma_{\text{H}_2}^{1.18 \pm 0.02}$ and $\Sigma_{\text{SFR}} = 10^{-1.78 \pm 0.05} \times \Sigma_{\text{H}_2}^{0.67 \pm 0.03}$, respectively. 
The contamination of AGN apparently increases the index $N$. The analysis of \citet{ken98} might have been contaminated with AGNs, which are typically found in galaxies with circumnuclear starbursts.

\subsection{Comparison with Previous Studies}
\begin{figure}[h]
\begin{center}
\includegraphics[width=10cm]{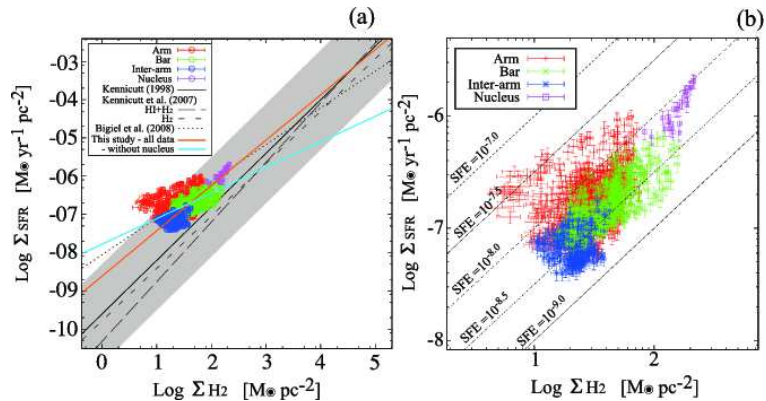}
\end{center}
\caption{
$\Sigma_{\text{H}_2}$ vs $\Sigma_{\text{SFR}}$ plot (Kennicutt-Schmidt plot). Left: $\Sigma_{\text{H}_2}$ vs $\Sigma_{\text{SFR}}$ at the resolution of 6$\arcsec$ ($\sim$ 500 pc) (color) at the scale of Kennicutt's Figure 9 (1998). Right: the same as the left panel, but at a larger scale.}
\label{fig:ks}
\end{figure}

Figure \ref{fig:ks} (a) compares our result with those from previous studies. The solid black line is the result of \citet{ken98}; $N$ = 1.4 $\pm$ 0.15. 
They calculated gas surface densities with CO data (gas density), and \ion{H}{1} and CO data (total gas density). The grey region indicates the scatter in Kennicutt (1998, their $Figure$ 9). 
Dashed lines are the fit in \citet{ken07}; the large and small dashes are the fits for \ion{H}{1} and CO ($N$ = 1.56 $\pm$ 0.04), and for CO-only ($N$ = 1.37 $\pm$ 0.03), respectively.
The dotted line is from \citet{big08}, and is for $N$ = 0.96 $\pm$ 0.07. Our result, $N$ = 0.67, is closer to \citet{big08}, than to \citet{ken98}, though it's smaller than the linear correlation $N$ =1. 

Our data lie roughly in the range of Kennicutt (1998, grey region), although some data points deviate from the exact range. The deviations occur on the spiral arms, where the offsets between the gas and star forming regions are large (\S 4.4). 
At the 500 pc resolution, the KS law is present, but suffers slightly from the offset effect. Two types of scatter appear in Figure \ref{fig:ks} (a), (b). One is the scatter of NGC 4303 itself, i.e., its deviation from the average of galaxies, and the other is within the disk of NGC 4303. 
The distribution of our points is concentrated in a small region, and the scatter within the galaxy is smaller than that of \citet{ken98}. Therefore, the scatter among galaxies seems dominant. 
Overall, NGC 4303 fits on the plot of \citet{ken98}, but has slightly higher SFE than the average.

Both $\Sigma_{\text{H}_2}$ and $\Sigma_{\text{SFR}}$ show one order of magnitude scatters in Figure \ref{fig:ks} (b), which is consistent with the results of Bigiel et al. (2008, their $Figure$ 4). Therefore, an order of magnitude scatter of star formation activity is common at a sub-kpc scale. \citet{ler08} showed that SFE varies strongly with local conditions. Since the KS law breaks down at the scale of a few 100 pc, we could attribute the variation of SFE to local environments at the scale of a few 100 pc. One of the key environmental factors might be galactic dynamics (e.g. spiral arms). 

We verify a scatter of NGC 4303 itself among nearby galaxies, i.e. a scatter from the KS law. Rough estimation of scatter shows that our results are a factor of $\sim$ 5 higher than the that of \citet{ken98}. However, the KS plot has a $\pm$ 1 order of magnitude scatter, as we discussed in \S 5.1. Even though our data are the fit by \citet{ken98}, they are still within the scatter (grey region).

\section{SUMMARY}
We observed the barred spiral galaxy \object{NGC 4303} in the $^{12}$CO(J=1-0) line with NRO45 and CARMA. The combination of NRO45 and CARMA provided an unprecedented high image fidelity as well as a high angular resolution (3$\farcs$2 $\sim$ 250 pc), which are critical for the accurate measurements of gas surface density and mass at high resolution. We discussed SFR and SFE quantitatively. Our results are summarized as follows: \\
\begin{enumerate}
\item CO emission is detected over the entire disk, i.e., almost everywhere including interarm regions and the downstream side of the bar. There are remarkable concentrations along the offset ridges of the bar and in the ring structure in the nucleus (r $\sim 1.6$ kpc area). The gas in the spiral arms extend from the end of the offset ridges toward the outer region. The surface densities in the outer spiral arms and offset ridges are similar at high resolution. 

\item Spatial offsets between H$\alpha$ and CO peaks exist along the spiral arms.  H$\alpha$ emission is seen at the downstream side of gas flow, while the CO emission is upstream of the gas flow. The delay of star formation from the formation of GMC on spiral arms would cause such offsets.

\item The azimuthal averaged SFE decreases steeply from the circumnuclear disk to the bar, and increases toward the spiral arms. The comparison of SFE in the bar and spiral arms shows that SFE is about twice as high in the arms as those of in the bar.

\item Extreme $\Sigma_\text{SFR}$ and SFE are found in the spiral arms, but not in the bar, indicating that the trigger of star formation is related not only to the amount of available gas, but also to the environment, such as galactic dynamics around spiral arms and the bar. The presence of the active star forming regions along the spiral arms confirms the visual impression that star formation is more active in spiral arms, or reduced significantly in bar.

\item The SFE derived with a metallicity-dependent $X_\text{CO}$ does not change the conclusion, i.e. higher SFE in the spiral arms than in the bar, since the difference between the bar and the spiral arms is reduced by only around 30$\%$. However, SFE in the circumnuclear regions is a factor of 2 -- 3 higher than the results with the SFE derived by a standard $X_\text{CO}$,  since metallicity of the circumnuclear is significantly high.

\item The KS law appears to break down at our highest spatial resolution ($\sim 250$ pc); due to the spatial offsets we find between parental gas and star forming regions. The correlation reappears if we smooth the images to $\sim 500$ pc resolution. The least squares fit of all data in the KS law is $\Sigma_{\text{SFR}} = 10^{-2.61 \pm 0.04} \times \Sigma_{\text{H}_2}^{1.18 \pm 0.02}$. NGC 4303 lies roughly in the range of \citet{ken98}, though it has slightly higher SFE than average value among nearby galaxies.

\end{enumerate}

\acknowledgments
We are very grateful to Yasutaka Kurono for helping us to combine our data, Takeshi Okuda for data analysis and discussions, Norio Ikeda for the 3-D FITS viewer package {\it ``FAZZ''\footnote{see http://hibari.isas.jaxa.jp/nikeda/fazz/fazz.html}} and Jennifer Donovan Meyer and James Barrett for helpful comments on the English.
We thank the NRO staff for NRO45 observations, the CARMA staff and CARMA summer school 2008 participants for CARMA observations and lectures, and helpful experience. This work made use of the Spitzer Space Telescope, which is operated by the Jet Propulsion Laboratory, California Institute of Technology under a contract with NASA, and the NASA/IPAC Extragalactic Database (NED). Support for CARMA construction was derived from the Gordon and Betty Moore Foundation, the Eileen and Kenneth Norris Foundation, the Caltech Associates, the states of California, Illinois, and Maryland, and the National Science Foundation. Ongoing CARMA development and operations  are supported by the National Science Foundation under a cooperative 
agreement, and by the CARMA partner universities. This research was partially supported by HST-AR-11261.01.

\vspace{1cm}

\begin{table}[h]
\begin{center}
\caption{Parameters of NGC 4303.}
\begin{tabular}{llcl}
\tableline\tableline
				Parameters & Value \\ 
\tableline				
				Name & \object{NGC 4303} / M 61 \\
				Hubble type\footnotemark{a} & SABbc \\
				V$_{lsr}$\footnotemark{b} &1556.5 km s$^{-1}$ \\
				Distance\footnotemark{c} & 16.1 Mpc \\
				Linear scale of 1$\arcsec$ & 78.1 pc \\
				Diameter\footnotemark{a} &  \\
				Major & 6$\farcm$5 \\
				Minor & 5$\farcm$8 \\
				Position Angle\footnotemark{d} & 312.2$\degr$ \\
				Inclination\footnotemark{d} & 27.8$\degr$ \\
				AGN type & Sy2 / \ion{H}{2} \\
\tableline
\end{tabular}
\footnotetext{a}{de Vaucouleurs et al. (1991)}
\footnotetext{b}{Koda $\&$ Sofue (2006)}
\footnotetext{c}{Ferrarese et al. (1996)}
\footnotetext{d}{This study.}
\label{table:4303}
\end{center}
\end{table}

\begin{table}[h]
\begin{center}
\caption{NRO45 and CARMA observational parameters.}
\begin{tabular}{lrrr}
\tableline\tableline
				Parameters & NRO45 & CARMA \\
\tableline
				Date & March 9th to 12th, 2008 & October to November, 2007, June, 2008 \\
				Field center (J2000) & 12$^{h}$21$^{m}$54.90$^{s}$, +4$^{d}$28$^{m}$25.1$^{s}$ & 12$^{h}$21$^{m}$54.9$^{s}$, +4$^{d}$28$^{m}$25.6$^{s}$  \\
				Observed Frequency & 114.67641 GHz & 114.67641 GHz \\
				Observed Region & 8$\arcmin$ $\times$ 8$\arcmin$ (37 kpc $\times$ 37 kpc) & 19 points mosaicing \\
				Bandwidth & 512 MHz & 3 $\times$ 62 MHz \\
\tableline
\end{tabular}
\end{center}
\end{table}

\begin{table}[h]
\begin{center}
\caption{NRO45, CARMA and combined data results.}
\begin{tabular}{lrrr}
\tableline\tableline
				Parameters & NRO45 & CARMA & Combined data  \\
\tableline
				Velocity resolution $\Delta V$ & 2.6 km s$^{-1}$ & 2.6 km s$^{-1}$ & 2.6 km s$^{-1}$  \\
				The rms noise in a channel & 73 mK & 43 mJy beam$^{-1}$ & 34 mJy beam$^{-1}$ \\
				The rms noise in an intensity map & 1.5 K km s$^{-1}$ & 810 mJy beam$^{-1}$ km s$^{-1}$ & 710 mJy beam$^{-1}$ km s$^{-1}$ \\
				 Beam / Synthesized beam size & 22$\farcs$1 & 3$\farcs$1 $\times$ 2$\farcs$5 & 3$\farcs$2 $\times$ 3$\farcs$2 \\
				Total flux & 2.8 $\times$ 10$^3$ Jy km s$^{-1}$ & 9.8 $\times$ 10$^2$ Jy km s$^{-1}$ & 2.9 $\times$ 10$^3$ Jy km s$^{-1}$ \\
\tableline
\end{tabular}
\label{table:comparison}
\end{center}
\end{table}

\begin{table}[h]
\begin{center}
\caption{The average of SFR and SFE over the entire disk regions, bar and spiral arms.}
\begin{tabular}{lrrr}
\tableline\tableline
		 & The entire disk & The bar & The spiral arms \\
\tableline
		SFR [M$_{\sun}$ yr$^{-1}$] & 8.3 $\times$ 10$^{-2}$ & 7.6 $\times$ 10$^{-2}$ & 9.8 $\times$ 10$^{-2}$ \\
		SFE [yr$^{-1}$] & 5.7 $\times$ 10$^{-9}$ & 4.2 $\times$ 10$^{-9}$ & 7.4 $\times$ 10$^{-9}$ \\
\tableline
\end{tabular}
\end{center}
\end{table}

\end{document}